\documentclass[journal=jctcce,manuscript=article]{achemso}

\usepackage[version=3]{mhchem} % Formula subscripts using \ce{}
\usepackage{physics}
\usepackage{braket}
\usepackage{mhchem}
\usepackage{color}

\author{Leon Otis}
\affiliation{Department of Physics, University of California Berkeley, CA 94720,USA }
\author{Isaac M. Craig}

\author{Eric Neuscamman}
\affiliation{Department of Chemistry, University of California Berkeley, CA 94720,USA}
\altaffiliation{Chemical Sciences Division, Lawrence Berkeley National Laboratory, Berkeley, CA, 
94720, USA}

\email{eneuscamman@berkeley.edu}

\title
  {A hybrid approach to excited-state-specific variational Monte Carlo
  and doubly excited states}

\begin{document}

% \begin{tocentry}

% Some journals require a graphical entry for the Table of Contents.
% This should be laid out ``print ready'' so that the sizing of the

% Inside the \texttt{tocentry} environment, the font used is Helvetica
% 8\,pt, as required by \emph{Journal of the American Chemical

% The surrounding frame is 9\,cm by 3.5\,cm, which is the maximum
% permitted for  \emph{Journal of the American Chemical Society}
% graphical table of content entries. The box will not resize if the
% content is too big: instead it will overflow the edge of the box.

% This box and the associated title will always be printed on a
% separate page at the end of the document.

% \end{tocentry}

\begin{abstract}
We extend our hybrid linear-method/accelerated-descent variational
Monte Carlo optimization approach to excited states and
investigate its efficacy in double excitations.
In addition to showing a superior statistical efficiency when
compared to the linear method, our tests on small molecules show good energetic agreement with benchmark methods.
We also demonstrate the ability to treat double excitations in systems
that are too large for a full treatment by
selected configuration interaction methods via
an application to 4-aminobenzonitrile.
Finally, we investigate the stability of state-specific variance
optimization against collapse to other states' variance minima
and find that symmetry, ansatz quality, and sample size
all have roles to play in achieving stability.
\end{abstract}

\section{Introduction}

Accurate predictions about doubly excited states remain a significant challenge
for electronic structure methods. 
Although they are rare in the low-lying spectra of simple organic molecules, states
with significant or near-total doubly excited character are not uncommon in
aromatic systems, other $\pi$-conjugated settings, and transition metal
compounds.
Their difficulty can be understood by considering three factors.
First, they tend to be strongly multi-reference in character, which frustrates
traditional weakly-correlated quantum chemistry approaches.
Second, predicting accurate excitation energies requires a method to also
capture weak correlation effects, and capturing both strong and weak correlation
in medium to large systems remains an open challenge in electronic structure,
despite the recent progress in smaller systems.\cite{Booth2009,Blunt2015,White1999,Chan2002,Kurashige2009,Holmes2016,Sharma2017,Abrams2005,Loos2018,Garniron2018,Loos2019}
Third, they are excited states, which remain harder to treat than ground states.
When faced with a doubly excited state, many excited state
approaches are either not appropriate or difficult to afford.
Time-dependent density functional theory (TDDFT), at least within the adiabatic 
approximation, is famously incapable of predicting double excitations,\cite{Levine2006}
while equation of motion coupled cluster with singles and doubles (EOM-CCSD)
is much less accurate in these states than in single excitations.
\cite{Hirata2000}
Higher-level coupled cluster is more effective, \cite{Watson2012}
but the $O(N^8)$ and $O(N^{10})$ cost scalings of EOM-CCSDT and EOM-CCSDTQ 
make them difficult to use outside of small molecules.
Similarly, selected configuration interaction (sCI) methods
can establish benchmark results for double excitations in small molecules,
\cite{Loos2019}
but their exponential scaling makes them difficult to extend to larger systems.
For these reasons, more traditional multi-reference methods like
CASPT2\cite{Andersson1990,Andersson1992}  have long been favored when dealing
with double excitations.\cite{SerranoAndres1993,SerranoAndres1994,Nakayama1998,Ostojic2001,Dallos2004,Schreiber2008,Silva-Junior2010,Duman2012,Wen2018,Rauer2016,BenAmor2017,Loos2019,Loos2020}  
These methods have their limitations as well --- intruder states, smaller
active space sizes than sCI, sometimes-sensitive state-averaging choices ---
and so the development of alternative high-accuracy approaches to double
excitations, especially ones that rely on very different approximations
that could cross-validate current methods' predictions, 
remains an important priority.

Quantum Monte Carlo (QMC) methods are, in principle, a promising alternative
for doubly excited states and for difficult excited states more generally.
Thanks to an ability to employ correlation factors and projector Monte Carlo
methods to impart weak correlation effects on top of modest determinant
expansions that capture the primary strong correlations, QMC approaches
offer one route towards an integrated treatment of weak and strong correlation.
Moreover, the ability of variational Monte Carlo (VMC) to work with excited states,
either in a state-averaged
\cite{Cordova2007,Filippi2009,guareschi2016,Cuzzocrea2020,Feldt2020}
or a state-specific
\cite{Umrigar1988,Zhao2016,Shea2017}
manner, offers a clear route to extending these advantages to
studies of doubly excited states.
%The variational (VMC) and diffusion (DMC) Monte Carlo methods can be employed with a wide 
%variety of functional forms for the wave function ansatz
%\cite{Foulkes2001,Feldt2020} and 
%an ongoing challenge in the QMC community has been the development of algorithms for robust 
%optimization of variational parameters in the presence of noise.
However, these advantages are only useful in practice if these sophisticated
wave function forms can be optimized successfully for excited states.
In ground states, recent years have seen significant improvements in the size
and complexity of wave functions that can be treated by VMC wave function
optimization,
\cite{Umrigar2007,Sorella2007,Neuscamman2012,Zhao2017,Schwarz2017,Sabzevari2018,Otis2019}
from the development of the linear method (LM) to the adoption of
accelerated descent (AD) methods and even the combination of the two.
Excited state wave function optimization is less well developed, and
variance-based state-specific approaches --- which are useful when
dealing with high-lying states \cite{Garner2020}
or in cases where large dipole changes
raise concerns for state-averaging \cite{Flores2019}
--- have been shown to face stability issues in some surprisingly
simple settings. \cite{Cuzzocrea2020}

%The study of excited states brings further complications to problem of optimization in 
%QMC.
%While the ground state (and excited states that are lowest in their symmetry class) can 
%be handled by a minimization of the energy, there are in general multiple 
%excited state variational principles available, including both 
%energy-based state-averaged\cite{Cordova2007,Filippi2009,Cuzzocrea2020} and 
%variance-based state-specific\cite{Zhao2016,Shea2017} methods.
%The state-averaged approach has been successfully applied in multiple 
%studies\cite{Cordova2007,Filippi2009,Cuzzocrea2020}, but may be less accurate in cases 
%where the excited state of interest is described by significantly different orbitals from
%the other states\cite{Feldt2020}, a problem which is common to state-averaging in other 
%methods such as CASSCF.\cite{Tran2019,Meyer2014}
%Variance-based optimization, which we shall consider in much greater detail, offers an 
%appealing way to avoid the limitations of state-averaging, but recent 
%work\cite{Cuzzocrea2020} has raised the possibility that targeting a specific excited 
%state may be too unstable to obtain reliable excitation energies in practice.
%Within this context, there is then a need for both further development of optimization 
%tools and examination of the efficacy of the variational principles they are applied to.

To help improve excited-state-specific wave function optimization
in VMC, this study extends our hybrid LM/AD
optimization approach \cite{Otis2019} to a variance-based
excited state objective function and
tests both its stability and its efficacy for double excitations.
As in the ground state, we find that AD provides useful support
to the Blocked LM, and that their combination is more statistically
efficient than the LM alone.
In terms of stability against collapse to other states' variance
minima, we find that the difficulty that AD faces in making
large changes to the wave function \cite{Otis2019}
actually prevents it from causing stability issues, but that the
LM part of the algorithm faces the same challenges as
have been seen \cite{Cuzzocrea2020}
in Newton optimizations of the variance.
We therefore focus our stability testing on the LM, and find
that achieving stable optimizations is greatly aided by
enforcing symmetries, improving the sophistication of the
Jastrow factor, and using a sufficiently large sample size to
avoid stochastically jumping out of a shallow minimum.
In terms of efficacy for double excitations, we find the hybrid
approach to be superior to the LM in all cases, usually in terms
of statistical efficiency, but in one case also in terms of
successfully finding the minimum at all.
Comparisons against benchmark methods confirm VMC's accuracy,
while a demonstration in a doubly excited state of
4-aminobenzonitrile shows that the approach can be expanded
well beyond the reach of sCI methods.

%We present an extension of the previously developed hybrid optimization 
%scheme\cite{Otis2019} to a state-specific functional and apply it to several doubly 
%excited states in test case molecules. 
%We also investigate the same cyanine dye system where Filippi and coworkers have identified 
%severe failures of variance-based variational principles\cite{Cuzzocrea2020} and find that
%our optimization approaches can be made stable.
%In our doubly excited state applications, we find that the hybrid method can achieve 
%excitation energies of comparable or superior accuracy compared to the LM
%and with greater statistical efficiency.
%These excitation energies can be obtained with relatively simple Multi-Slater Jastrow (MSJ) 
%wave functions and are in good agreement with benchmark values.

\section{Theory}

\subsection{Excited-State-Specific Variational Monte Carlo}

While VMC has historically been used for the study of ground states, multiple functionals 
have been developed for the targeting and variational optimization of excited states.
We focus on the recently developed functional\cite{Zhao2016}, 
\begin{equation}
\label{eqn:targetAbove}
    \Omega (\Psi) = \frac{\Braket{\Psi | (\omega - H) | \Psi}}{\Braket{\Psi | (\omega - H)^2 | \Psi}}
\end{equation}
which is minimized when $\Psi$ is the eigenstate of lowest energy greater than $\omega$.
The formulation of VMC for this excited state functional proceeds analogously to the ground state case.
Just as ground state VMC seeks to minimize the expectation value of the energy, written as
\begin{equation}
\label{eqn:energy}
    E(\Psi) = \frac{\Braket{\Psi | H | \Psi}}{\Braket{\Psi | \Psi}}
    = \frac{\int d \mathbf{R}\Psi (\mathbf{R}) H \Psi(\mathbf{R})}{\int d \mathbf{R}\Psi (\mathbf{R})^2}
    = \frac{\int d \mathbf{R}\Psi (\mathbf{R})^2 E_L (\mathbf{R})}{\int d \mathbf{R}\Psi (\mathbf{R})^2} = \int d \mathbf{R} \rho (\mathbf{R})E_L (\mathbf{R})
\end{equation}
using the local energy $E_L (\mathbf{R}) = \frac{H \Psi (\mathbf{R})}{\Psi (\mathbf{R})}$ and probability density $\rho (\mathbf{R}) = \frac{\Psi(\mathbf{R})^2}{\int d \mathbf{R} \Psi (\mathbf{R})^2}$, we can write $\Omega$ as
\begin{equation}
\label{eqn:targetAboveDetail}
    \Omega(\Psi) = \frac{\int d \mathbf{R}\Psi (\mathbf{R}) (\omega - H) \Psi(\mathbf{R})}{\int d \mathbf{R}\Psi (\mathbf{R}) (\omega - H)^2 \Psi (\mathbf{R})}
    = \frac{\int d \mathbf{R} \rho (\mathbf{R})(\omega - E_L (\mathbf{R}))}{\int d \mathbf{R}\rho (\mathbf{R}) (\omega - E_L (\mathbf{R}))^2}. 
\end{equation}
However, while the probability distribution $\rho$ has a useful zero-variance
property\cite{Assaraf1999}, it may be less useful than other importance sampling functions 
when estimating quantities such as the energy variance and matrix elements used in the
optimization algorithms for minimizing $\Omega$. \cite{Trail2008a,Trail2008b,robinson2017vm,Flores2019}
Within this work, we use the importance sampling function
\begin{equation}
\label{eqn:guiding}
    |\Phi|^2 = |\Psi|^2 + c_1\sum_{i} |\Psi^i|^2 + c_2\sum_{j}|\Psi^j|^2 + c_3\sum_{k}|\Psi^k|^2
\end{equation}
where $c_1$,$c_2$,$c_3$ are weights on sums of squares of wave function derivatives
$\Psi^i, \Psi^j,\Psi^k$ for Jastrow, CI, and orbital parameters, respectively.
Effective choices for $(c_1,c_2,c_3)$ may be system-dependent and require some 
experimentation in practice.
In our results, we use $(0.0,0.0001,0.0)$ for our stability tests on a model 
cyanine dye, $(0.0004,0.0002,0.0)$ for the carbon dimer, $(0.0001,0.0001,0.0)$ for nitroxyl, 
glyoxal, and acrolein, and $(0.0001,0.0,0.0)$ for cyclopentadiene and 4-aminobenzonitrile.
We have found these values enabled optimization to lower target function values
than we could achieve using $|\Psi|^2$ and that using nonzero $c_3$ for the orbital parameter 
derivatives also resulted in poorer target function values. 
However, our exploration of these importance function parameters was not exhaustive and
different values may also result in equally good target function values.
The general intuition is to include wave function derivative terms, as importance sampling a linear combination of the current wave function probability density and that of its parameter derivatives allows us to obtain a better statistical estimate of the LM Hamiltonian. Whenever the parameters have large effects on the nodal surface, importance sampling $|\Psi|^2$ will lead to a large statistical uncertainty in this Hamiltonian. Importance sampling $|\Phi|^2$ allows us to better sample areas of parameter space where the current $\Psi$ has negligible probability density, but linear updates to $\Psi$ may not, obtaining a better Hamiltonian estimate and therefore better parameter updates in the process. Because this guiding function also has considerably fewer nodes, it can further help avoid the divergence in the variance of the 
variance\cite{Trail2008a,Trail2008b} encountered with 
$|\Psi|^2$ while also keeping the distribution close to $|\Psi|^2$.\cite{robinson2017vm}
%We note that this is the primary effect when including only the derivatives of parameters which cannot directly effect the nodal surface, such as the Jastrow parameters.
%\textbf{End of area Isabel modified.}
We also employ clipping of the samples based on the 
value of the local energy\cite{Umrigar1993,Pfau2019}.
We compute a deviation of the form $\frac{1}{N}\sum |E_L(\mathbf{R}) - \bar{E}|$ 
where $\bar{E}$ is the mean local energy of all initial N samples.
Samples with local energies more than 5 times the deviation from $\bar{E}$ are discarded
and only the remainder are used in computations for the VMC optimization.
We find that clipping can improve optimization performance, especially as more flexible 
ansatzes with larger numbers of parameters are considered, by reducing the statistical 
uncertainty in the matrix elements and derivatives used by the algorithms.
Essentially, this guards against large parameter updates.

The use of $\Omega$ for excited states also requires careful handling of the input $\omega$.
For a generic fixed choice of $\omega$, $\Omega$ will not be size-consistent and so $\omega$ 
must be updated to transform $\Omega$ into state-specific variance minimization, which is 
size-consistent.\cite{Shea2017}
There are multiple strategies for achieving this transformation to have both state-specificity
and size-consistency, such as a linear interpolation between the initial fixed value of 
$\omega$ and the floating value of $E-\sigma$,\cite{Shea2017,Cuzzocrea2020} or, as in 
this work, a series of fixed-$\omega$ optimizations with $\omega$ updated between each one 
until self-consistency between it and $E-\sigma$ is reached.
The details of how $\omega$ is varied in a calculation is one potential source of instability 
in the optimization as the target function being minimized with respect to wave function 
parameters is now changing and there is the possibility of slipping outside the basin of
convergence, particularly if $\omega$ is varied rapidly.

\subsection{Linear Method}

The LM\cite{Nightingale2001,Umrigar2007,Toulouse2007,Toulouse2008} is based on a first order Taylor expansion of the wave function

\begin{equation}
\label{eqn:lmTaylor}
    \Psi (\mathbf{p}) = \Psi_0 + \sum_i \Delta p_i \Psi_i
\end{equation}
using $\Psi_i = \frac{\partial \Psi(\mathbf{p})}{\partial p_i}$ for first order parameter
derivatives of the wave function and $\Psi_0$ for the wave function at the current parameter
values $\mathbf{p}$.
Seeking to minimize the target function $\Omega$ with respect to $\mathbf{p}$ leads to the 
generalized eigenvalue problem

\begin{equation}
\label{eqn:lmEigen}
    (\omega - \mathbf{H}) \hspace{0.6mm} \mathbf{c}
    = \lambda \hspace{0.6mm} (\omega - \mathbf{H})^2 \hspace{0.6mm} \mathbf{c}
\end{equation}
 which can be solved to produce an update vector $\mathbf{c} = (1,\Delta \mathbf{p})$.
 The matrices are constructed in the basis of the initial wave function and its first order
 parameter derivatives, so we have matrix elements of the form
 \begin{equation}
 \label{eqn:lmLHS}
     \Braket{\Psi_i | \omega - H | \Psi_j}
 \end{equation}
 and
 \begin{equation}
 \label{eqn:lmRHS}
     \Braket{\Psi_i | (\omega - H)^2 | \Psi_j}.
 \end{equation}
These matrix elements are evaluated within VMC using parameter derivatives of the wave function
and the local energy, and we have found that employing the modified guiding function $|\Phi|^2$
can be crucial for obtaining accurate estimates and effective LM optimizations.
 In practice, the Hamiltonian matrix of the LM is modified with the addition of shift 
 matrices\cite{Kim2018,Otis2019} which help prevent incautiously large parameter changes in 
 the optimization and can noticeably influence the LM's stability and performance.
Our implementation of the LM uses by default an adaptive scheme where 
three sets of shift values are used to produce candidate parameter updates and a correlated 
sampling procedure is used to determine which one, if any, should be taken to improve the 
target function value.
This approach allows the value of the shifts to vary over the course of the optimization, with 
the aim of allowing the LM to safely take large steps in parameter space early on while 
automatically becoming more cautious when close to the minimum, where statistical uncertainty 
will eventually prevent steps from being able to resolve downhill.
 
Much recent work on VMC for excited 
states\cite{Zhao2016,robinson2017vm,Flores2019,Zhao2019,Garner2020,Zimmerman2009,Filippi2009,Send2011}
has relied on the LM for the task of wave function optimization.
However, the LM has multiple limitations, particularly a memory cost that increases with 
the square of the number of optimizable parameters, and applications with more than 10,000
parameters are rare.
The quadratic growth in the number of LM matrix elements exacerbates the nonlinear bias of 
the LM and eventually leads to the underdetermined regime where there are too few samples to 
effectively estimate the LM matrices and the step uncertainty is large.
To help address these issues, a variant of the LM known as Blocked LM has been
developed\cite{Zhao2017} that divides the parameter set into $N_b$ blocks and performs a 
LM-style matrix diagonalization for each block.
Some number $N_k$ of the eigenvectors from each block are retained and combined together along 
with $N_o$ other directions in parameter space for a second LM diagonalization to obtain 
an update in the full parameter space.
The lower dimension of the matrices constructed by the Blocked LM alleviates the issues faced
by the standard LM at the cost of having to run over the samples twice instead of only once.
Further details on the Blocked LM can be found in the original publication\cite{Zhao2017}
and we describe our choices on the number of blocks and retained directions in the
supplementary material.
A recent study\cite{Otis2019} of optimization methods for ground state optimization indicates 
that tighter and more statistically efficient convergence can be obtained through a hybrid 
combination of the Blocked LM and AD than when using the LM alone.
 
\subsection{Hybrid Optimization}

There are multiple flavors of AD optimization methods that require only first order 
parameter derivatives.
These methods are widespread in the machine learning community and have been increasingly
used in the context of VMC.\cite{Schwarz2017,Sabzevari2018,Luo2018,Mahajan2019,Otis2019}
They offer some appealing advantages compared to the LM, including a memory cost linear in 
parameter number, a lower per-iteration sampling cost to estimate derivatives, and a reduced 
nonlinear bias from the stochastic evaluation of those derivatives.
However, comparisons\cite{Otis2019} between AD methods and the LM indicate that the former 
may struggle to reach the minimum in parameter space at comparable computational effort,
especially when the wave function contains many challenging nonlinear parameters.

An alternative approach that shows the potential to benefit from the strengths of both
classes of methods is to take a hybrid combination that alternates between them.\cite{Otis2019}
Sections of gradient descent optimization can be used to identify important directions in 
parameter space, which can be provided as input to the Blocked LM to help it account for 
coupling between blocks of parameters.
The inclusion of Blocked LM steps allows the hybrid method to more efficiently reach the 
vicinity of the minimum, while the sections of AD enable tighter convergence by correcting
poor parameter updates by the Blocked LM due to step uncertainty.
In addition, the use of AD and Blocked LM enables the optimization of larger parameter sets 
that are beyond the reach of the standard LM.
In energy minimization we found it especially useful to follow the hybrid optimization with a
final section of optimization using only descent, which has been found to more efficiently 
achieve lower final statistical uncertainties than optimization based solely on the 
LM.\cite{Otis2019}

The hybrid scheme can be used with any of the variety of AD methods, but in this work, 
we use a combination of Nesterov momentum with the RMSprop algorithm\cite{Schwarz2017} that 
was found to work well for energy minimization.
It is specified by the following recurrence relations.

\begin{equation}
\label{eqn:rmspropMomentum}
    p_i^{k+1} = (1-\gamma_k e^{-(\frac{1}{d})(k-1)})q_i^{k+1} - \gamma_k e^{-(\frac{1}{d})(k-1)} q_i^k
\end{equation}
\begin{equation}
\label{eqn:rmspropUpdate}
   q_i^{k+1} = p_i^k - \tau_k \frac{\partial \Omega (\mathbf{p})}{\partial p_i}
\end{equation}
\begin{equation}
\label{eqn:rmspropRecur}
\lambda_0=0 \hspace{7mm}
\lambda_k = \frac{1}{2} + \frac{1}{2}\sqrt{1+4\lambda_{k-1}^2} \hspace{7mm}
\gamma_k = \frac{1-\lambda_k}{\lambda_{k+1}}
\end{equation}
\begin{equation}
\label{eqn:rmspropStep}
    \tau_k = \frac{\eta}{\sqrt{E[(\frac{\partial\Omega}{\partial p_i})^2]^{(k)}} + \epsilon}
\end{equation}
\begin{equation}
\label{eqn:rmspropAvg}
    E[(\partial \Omega)^2]^{(k)} = \rho E\left[\left(\frac{\partial \Omega}{\partial p_i}\right)^2\right]^{(k-1)} + (1-\rho)\left(\frac{\partial\Omega}{\partial p_i}\right)^2
\end{equation}
Equations \ref{eqn:rmspropMomentum} through \ref{eqn:rmspropRecur} describe how a parameter 
$p_i^k$ of the wave function on step $k$ of the optimization is updated using knowledge of
both the current and previous values of the target function derivatives.
The step size $\tau_k$ is adaptively adjusted using a running average of the square of 
parameter derivatives 
according to equations \ref{eqn:rmspropStep} and \ref{eqn:rmspropAvg}, where $\rho$ sets the 
relative weighting between the past average $E\left[\left(\frac{\partial \Omega}{\partial p_i}\right)^2\right]^{(k-1)}$ and 
the current value $\left(\frac{\partial\Omega}{\partial p_i}\right)^2$.
The quantities $d$,$\eta$, $\rho$, and $\epsilon$ are hyperparameters whose values are chosen 
by the algorithm's user.
For all our results, we have used $d=100$, $\rho = 0.9$ and $\epsilon = 10^{-8}$.
The value of $\eta$ sets an overall scale for step size and can have a significant influence on
optimization performance depending on the choices made for different types of parameters.
Our choices for $\eta$ are discussed in the supplementary material.

\subsection{Optimization Stability}

When working with an exact ansatz and an infinite sample size,
any non-degenerate Hamiltonian eigenstate possesses its own
variance minimum, \cite{Umrigar1988}
and an optimization with a guess sufficiently close to that
minimum will be stable.
In practice, sample sizes are finite, and the use of an approximate
ansatz may lead some states' variance minima to be artificially
shallow or to disappear entirely, and so there is a real
possibility that state-specific variance minimization
will be unstable. \cite{Cuzzocrea2020}
If the minimum has indeed disappeared, then the only remedy is
to improve the ansatz quality, as no amount of statistical
precision will allow an optimization to find a minima that is
not present.
On the bright side, advances in VMC trial 
functions\cite{Filippi1996,Drummond2004,Goetz2017,Goetz2019,Filippi2016,Assaraf2017,Casula2003,Casula2004,Marchi2009}
offer a strong toolkit for improving ansatz quality, albeit one
whose use can increase optimization difficulty by increasing the
number of variational parameters.
If anything, this reality further motivates the development
of optimizers, like the hybrid approach tested here, that are
designed to handle large, challenging trial functions.
In Section \ref{sec::stability}, we will explore the issue
of optimization stability by starting with a very simple ansatz
for which some states
lack variance minima and then making improvements, either by
enforcing symmetries or by enhancing the ansatz in order
to make the variance minima re-emerge.

If the variance minimum is present but shallow, either due to
an approximate ansatz or simply to a near-degeneracy in the
spectrum, then the nature of the optimization update steps, and
especially their statistical uncertainty,
will determine whether an optimization is stable.
Strictly speaking, finite-sample-size VMC optimizations always
contain at least some risk of selecting a step in the tail of
the distribution of steps that, if taken, will move the
optimization out of the basin of convergence for one variance
minimum and into the basin for another.
To help guard against this issue, a common practice in VMC,
and one that we do use for our LM steps,
is to verify on a new independent sample (usually using
correlated sampling) that the step about to be taken does indeed
lower the objective function.
Even without statistical uncertainty, step size control is
important in nonlinear optimization, an issue that
is typically addressed by trust radius methods that guard
against single steps making overly large parameter changes.
\cite{Sorensen1982}
In the LM, one approach to this issue is a diagonal shift,
\cite{Umrigar2007}
and our LM implementation employs both this
shift and an overlap-based shift \cite{Kim2018}
for these purposes.
In Section \ref{sec::stability}, we will see an example of
an optimization that only becomes stable with a large enough
sample size due to a shallow minimum created by a near-degeneracy.

\subsection{Wave Functions}

For demonstrating the hybrid method's effectiveness within our overall QMC methodology for 
excited states, we consider multiple types of parameters in our trial wave functions.
Our ansatz is the Multi-Slater Jastrow wave function, which has the form 

\begin{equation}
\label{eqn:psi}
    \Psi = \psi_{MS} \psi_J 
\end{equation}
\begin{equation}
\label{eqn:psiMS}
    \psi_{MS} = \sum_{i=0}^{N_D} c_i D_i
\end{equation}
\begin{equation}
\label{eqn:psiJ}
    \psi_J = \exp{\sum_i \sum_j \chi_k(|r_i - R_j|) + \sum_k \sum_{l>k} u_{kl} (|r_k - r_l|)}
\end{equation}
 where $D_i$ are Slater determinants with coefficients $c_i$ and the Jastrow factor $\psi_J$ is
constructed from splines that make up the functions $\chi_k$ and $u_{kl}$ for the
respective one- and two-body terms.\cite{Kim2018}
The MSJ ansatz is a common choice in QMC, but it can be augmented further to describe more 
correlation at the price of a more challenging optimization.
Two means for doing so are to add a more complicated Jastrow factor and to 
optimize molecular orbital shapes.

A variety\cite{Umrigar1988,Drummond2004,LopezRios2012,Luchow2015,Huang1997,Casula2003,Casula2004,Sterpone2008,Beaudet2008,Marchi2009,Zen2015} of many-body Jastrow factors have been considered for improving QMC ansatzes, but 
in this work we will limit ourselves to adding a three-body term and a
number-counting factor.
The three-body term is constructed from polynomials of interparticle distances and can be
written in the following form.\cite{Drummond2004}
\begin{equation}
\label{eqn:threeBodyJastrow}
    u(r_{\sigma I},r_{\sigma' I},r_{\sigma \sigma'}) = \sum_{l=0}^{M_{eI}} \sum_{m=0}^{M_{eI}} \sum_{n=0}^{M_{ee}} \gamma_{lmn} \, r_{\sigma I}^l r_{\sigma' I}^m r_{\sigma \sigma'}^n \, (r_{\sigma I} - \frac{r_c}{2})^3 (r_{\sigma' I} - \frac{r_c}{2})^3 \, \Theta (r_{\sigma I} - \frac{r_c}{2}) \Theta (r_{\sigma' I} - \frac{r_c}{2})
\end{equation}
The maximum polynomial orders are set by $M_{eI}$ and $M_{ee}$ for the electron-ion and 
electron-electron distances respectively. The $\gamma_{lmn}$ are the set of optimizable 
parameters in this polynomial and are subject to constraints for ensuring the Jastrow satisfies
symmetry under exchange and cusp conditions, which can be found in the original 
publication.\cite{Drummond2004}
Finally, the Theta functions require the three-body term become zero for electron-ion 
distances more than half a chosen cutoff distance $r_c$ (10 bohr in our case).

Number-counting Jastrow factors are a recently developed\cite{Goetz2017,Goetz2019} ansatz
component and can be thought of as a many-body Jastrow factor in real space that aims to 
recover both strong and weak correlation.
They are based on a Voronoi partitioning of space, where the population of electrons in 
each region, $N_I$, is given by a sum of local counting functions $C_I$ at each electron coordinate.

\begin{equation}
\label{eqn:ncjfPop}
    N_I = \sum_i C_I (\mathbf{r_i}) = \sum_i \frac{g_I(\mathbf{r})}{\sum_j g_j (\mathbf{r})}
\end{equation}
where 
\begin{equation}
\label{eqn:ncjfGauss}
    g_j(\mathbf{r}) = \exp ((\mathbf{r} - \boldsymbol{\mu})^T \mathbf{A} (\mathbf{r} - \boldsymbol{\mu}) + K)
\end{equation}
are Gaussian basis functions about a center $\boldsymbol{\mu}$.
With these populations $N_I$, we can construct the Jastrow factor
\begin{equation}
\label{eqn:psiNCJF}
    \psi_C = \exp(\sum_{IJ} F_{IJ} N_I N_J + \sum_K G_K N_K) 
\end{equation}
which can be tacked on to our overall expression for $\Psi$.
The $F_{IJ}$ and $G_K$ are variational parameters though the latter is in practice
eliminated with a basis transformation of the region populations and so we only optimize $F_{IJ}$.\cite{Goetz2019}

State-specific orbital optimization is useful for obtaining accurate VMC results on 
particular excited state phenomena including charge transfer\cite{Flores2019} and core 
excitations\cite{Garner2020} and can avoid some of the pitfalls of state-averaged approaches.\cite{Tran2019}
Recent work\cite{Filippi2016,Assaraf2017} with the table method has enabled the efficient 
calculation of orbital parameter derivatives in MSJ wave functions even for large expansions 
and we refer the reader to the original publications for details.
However, we do note that, despite these advances, orbital optimization remains the most 
challenging part of the optimization, likely due to its inherently high degree of nonlinearity
and the fact that it alters the nodal surface.
One potential alternative would be to obtain state-specific orbitals from another 
method\cite{Tran2019} while optimizing only Slater coefficients and Jastrow factors, and we
make a preliminary exploration of this idea in our results.

\subsection{Variance Matching}

While optimization of the parameters improves the absolute quality of the wave functions, the
results are still approximate due to the limited ansatz and accurate determination of excitation energy differences 
requires cancellation of errors.
This relies on a balanced treatment of both the ground and the excited state, which we 
attempt to obtain using variance matching.
As shown in previous work,\cite{robinson2017vm,Flores2019} this approach improves predicted
excitation energies by optimizing ansatzes of different CI expansion lengths for the two 
states so that their variances are approximately equal.
To facilitate interpolation, the variances for a series of excited state calculations at 
different expansion lengths can be fit to an analytic form such as the power law decay
\begin{equation}
    \sigma^2(N) = c + \frac{d}{N^{\alpha}}
\end{equation}
to determine parameters, $c$, $d$, and $\alpha$.
For a given expansion length for the ground state and a resulting variance $\sigma^2_g$,
we can estimate the expansion length $N^*$ that will yield a matching variance for the 
excited state and take the corresponding energy when computing our prediction for the 
excitation energy.\cite{robinson2017vm}
In practice, some additional varying of $N$ by hand can be performed to find an explicit 
variance match.
All our reported excitation energies were obtained using this explicit variance matching 
procedure.

\section{Results}

\subsection{Computational Details}
All our VMC calculations used an implementation of the described optimization algorithms within a development version of 
QMCPACK.\cite{Kim2018,Kent2020}
A recently developed set of pseudopotentials\cite{Bennett2017} and associated basis sets were 
used for all molecules.
For constructing our ansatzes, we have employed Molpro\cite{MOLPRO_brief} and 
PySCF\cite{Sun2018} for CASSCF calculations to generate Slater determinant expansions.
In one case, we instead use CASSCF to provide orbitals for a selected CI calculation in 
Dice.\cite{Holmes2016,Sharma2017}
CASPT2 calculations in Molpro\cite{MOLPRO_brief} were used alongside other methods' literature 
values in benchmarking our VMC results.
Specific active space and basis set choices are given in each system's section.
Molecular geometries, absolute energies for our doubly excited state calculations, and additional optimization details can be found in the supplementary material.

 \begin{figure}[H]
 \includegraphics[width=12cm]{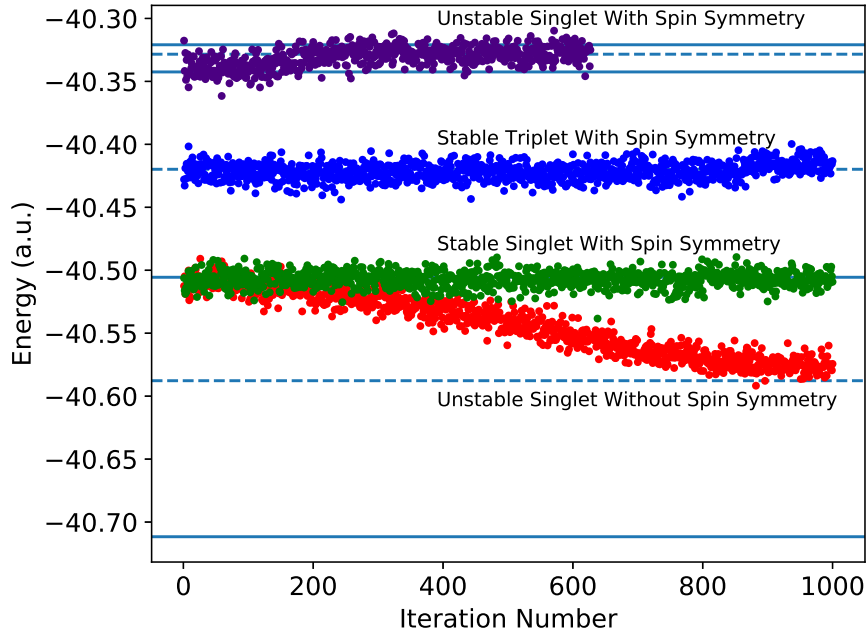}
 \caption{
 Examples of unstable and stable LM optimizations of $\Omega$ for CN5. 
We consider the singlet near -40.5 E$_h$, which is dominated by a HOMO to LUMO 
 excitation, as well as the next triplet and singlet above it, which are dominated by a HOMO-2 
 to LUMO excitation. 
 Initial wave function guesses were obtained from a diagonalization of the Hamiltonian in a
 26 determinant space with pre-optimized one- and two-body Jastrow factors present.
These Jastrows were optimized further along with the 26 determinant coefficients.
For each state, the value of $\omega$ was set to the E-$\sigma$ value from a one million sample evaluation using the initial wave function and held fixed throughout the optimization.
The LM shifts were kept constant at 0.1 for the diagonal shift and 1 for the overlap-based shift throughout the optimizations with 50,000 samples per iteration in all cases.
For the unstable cases, we found that using increased sampling effort alone did not ensure stability.
A proposed parameter update was rejected if our correlated sampling assessment predicted it 
would raise the target function value, but all optimizations contain hundreds of accepted 
steps.
Horizontal lines show the lowest 7 eigenenergies from our diagonalization, with solid lines for
singlet states and dashed for triplets.
 }
\label{fig:cn5_lm_results}
\end{figure}

\subsection{Stability in a Model Cyanine Dye}
\label{sec::stability}

To explore the stability of $\Omega$-based variance minimization,
we have performed a series of LM tests on the model cyanine dye
\ce{C3H3(NH2)2+} (denoted hereafter as CN5)
in which Filippi and coworkers discovered variance
optimization instabilities for some choices of the
trial function. \cite{Cuzzocrea2020}
In particular, they showed that while a CAS(6e,5o) wave function was stable, other 
ansatze from CAS(6e,10o) and CIPSI were not.
They also showed that a small CSF expansion, constructed of HOMO-to-virtual excitations
of B$_1$ symmetry, was especially prone to optimize to a
different state than the one targeted by the initial guess due to the absence of variance 
minima in this simpler case.
Here we perform similar tests to study the absence of variance minima, 
confirming that instabilities
are present when the sample size is small and the trial function
is simple, but also demonstrating that improvements in the
trial function and sampling effort can overcome these
instabilities.
To start, we note that virtual orbitals outside the CASSCF
active space are often not physical in their shapes, and so
for the states we target, we make sure that the primary
orbitals involved are within the (6e,5o) CASSCF active space.
In this active space, we performed an equal-weight
state-averaged CASSCF optimization of the lowest four B$_1$
singlet states using an aug-cc-pVDZ basis set in Molpro\cite{MOLPRO_brief}, after which we 
imported all determinants with
weights above 0.05 in any state into our VMC ansatz.
This resulted in a 26-determinant ansatz, or, in the cases
where we enforced singlet spin symmetry in VMC, a 13-CSF ansatz.
Note that this trial function differs from the one used previously,
\cite{Cuzzocrea2020} which is intentional, as the previous
approach examined some states whose dominant orbitals
were virtual in the quantum chemistry calculations and so may
not have been as well optimized as orbitals containing electrons.
Here, we try to ensure that orbital quality is balanced between
states by ensuring that the dominant orbitals of the states we
test are within our CASSCF active space.
Nonetheless, we can still find optimization
instabilities when using this simple ansatz, which confirms
that ansatz components beyond a small determinant expansion
can be necessary to achieve stable variance optimization.

As seen in Figure \ref{fig:cn5_lm_results}, the stability of
an optimization using our 26-determinant ansatz
%(which also includes simple 1- and 2-body Jastrow factors)
depends on which state is being targeted and in some cases
on whether or not spin symmetry is enforced.
First, consider the optimization that guesses the singlet
near -40.5 E$_h$ and then collapses to the triplet below it.
This failure shows that, at least at this
level of statistical resolution,
the singlet in question lacks a local variance minimum
in the variable space of the 26 different determinant
coefficients.
However, after enforcing spin symmetry by instead optimizing
the 13 coefficients for the singlet CSFs, the variance
minimum for this state reappears.
Next, we note that when we 
optimize the 26 determinant coefficients starting with a
guess for the triplet just below -40.4 E$_h$, the
optimization is stable.
Thus, in these cases, moving to a more sophisticated ansatz is
not necessary so long as symmetries are enforced.
The same cannot be said for the most difficult case we
consider, in which a guess for the singlet just above
-40.35 E$_h$ is unstable and optimizes to a higher singlet
even when spin symmetry is enforced.
With no more symmetries to make use of, we must conclude that
variance minimization is not stable for this state when using
this simple ansatz.

\begin{figure}[t]
 \includegraphics[width=12cm]{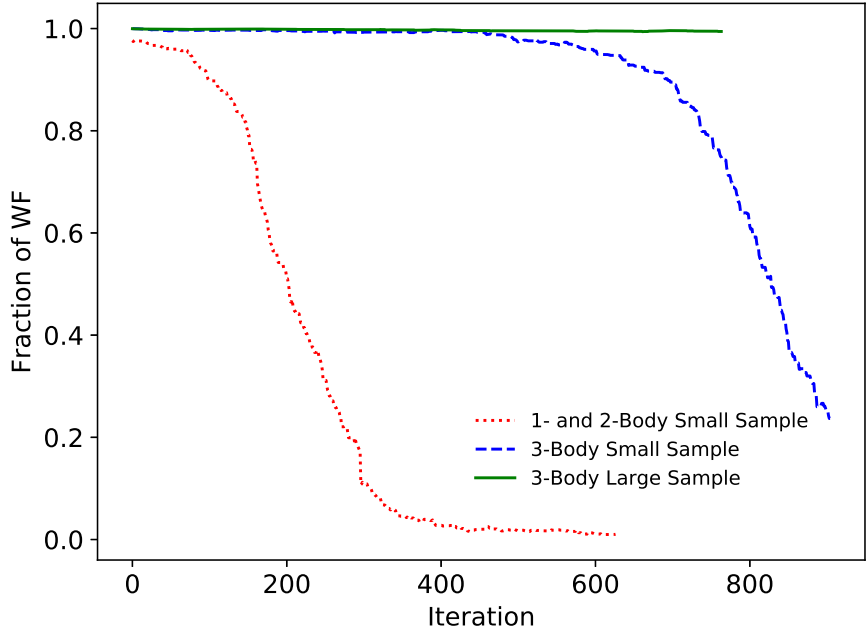}
 \caption{
 Fractions of the CI expansion corresponding to the third singlet's 
 dominant CSF over the course of LM optimizations. 
When starting from the energy eigenstate with only 1- and 2-body Jastrows are present, the weight of this CSF quickly collapses as 
the optimization drifts to a higher singlet.
 This optimization used 50,000 samples per iteration and starting from solely the dominant CSF  with an additional 3-body Jastrow delayed,
 but did not prevent, the drift at this sampling effort.
 Upon increasing the sample size to 500,000 samples per iteration, the 3-body Jastrow wave function optimization became stable, as shown by the solid line.
 Note that the CI expansion fraction is simply the sum of the squares of the determinant coefficients within the given CSF divided by the sum of all the squared determinant coefficients.} 
 \label{fig:cn5_lm_ci_stablity}
\end{figure}

In principle, variance minimization should become stable as
the ansatz is improved and the sample size increased.
The question, of course, is whether stability can be achieved
in practice, especially given that overly aggressive embellishments
of the ansatz may lead to an impossibly difficult optimization.
First, let us improve the ansatz by adding the 3-body
Jastrow factor introduced by Needs and coworkers\cite{Drummond2004},
which we first optimize for the guessed CSF expansion
with the CSF coefficients held fixed before finally optimizing
all variables together.
As seen in Figure \ref{fig:cn5_lm_ci_stablity}, which shows the key
CSF weights during the final optimization in which all parameters
are free to vary, this ansatz improvement
does lead to a stable optimization, but only when a larger sample
size is used, suggesting that the variance minimum is now present
but shallow.
Given how close in energy this state is to the singlet state
above it (about 15 mE$_h$),
a shallow minimum makes sense, which is a good reminder
that optimization objective functions based on the energy and
variance are at a disadvantage when states are nearly degenerate.
For low-lying states, state-averaged energy minimization can often deal with this type of 
situation, \cite{Cuzzocrea2020} but near-degeneracies remain a challenge for higher-lying states
or in situations where state-averaging introduces its own challenges.
In future, it may therefore be worth considering ways to involve other
properties in the objective function, possibly via a VMC
analogue of a generalized variational principle.
\cite{Shea2020}
Even with this near-degeneracy, which we stress involves states
significantly closer together in energy than any of those 
tested in the small-determinant-expansion case in the previous
study, \cite{Cuzzocrea2020} we do see that a modest improvement in
ansatz quality produces a stable $\Omega$-based LM optimization.

While this finding is reassuring, we must stress that difficult
optimization cases remain a serious challenge.
For example, Filippi and coworkers found \cite{Cuzzocrea2020}
that highly sophisticated wave functions derived from a (6,10)
active space faced optimization instabilities even though
wave functions derived from a smaller (6,5) active space
did not, showing that in practice it is not always
easy to predict when these instabilities will arise.
This in mind, we have checked carefully for signs of
optimization instability in the double excitations we now
turn to, and while we did not observe any such issues in these
states, finding a more complete resolution to this problem
is clearly an important direction for future research.

\subsection{Carbon Dimer}
For our doubly excited state applications, we first consider the carbon dimer, a very heavily 
studied system for the testing and development of theoretical
methods\cite{Abrams2004,Blunt2015,Booth2011,Loos2019}.
We use the $ 2 \; {}^1 \Sigma_g$ state, which is characterized by a HOMO to LUMO squared double
excitation, as a simple starting test case for validating the 
hybrid method's results against the LM's and assessing the accuracy of VMC.
In this case, selected CI methods are able to achieve millions of determinants and provide 
high quality benchmark data\cite{Holmes2017,Loos2019} on the vertical excitation energy that we
can use to assess our results with more compact wave functions.
We consider only an equilibrium bond length of 1.248 \r{A}.

To construct our ansatzes, we use a (8e,8o) CASSCF calculation in Molpro\cite{MOLPRO_brief}
with a carbon pseudopotential and the corresponding cc-pVTZ
basis.\cite{Bennett2017}
The resulting CI expansions are used in our variance matching procedure and we consider 
cases both for standard MSJ ansatzes using only one- and two-body Jastrows, and with an additional NCJF and orbital optimization.
The NCJF was produced using a set of 16 counting regions composed of 8 octants for each 
carbon atom.\cite{Goetz2019}

\begin{figure}[H]
\includegraphics{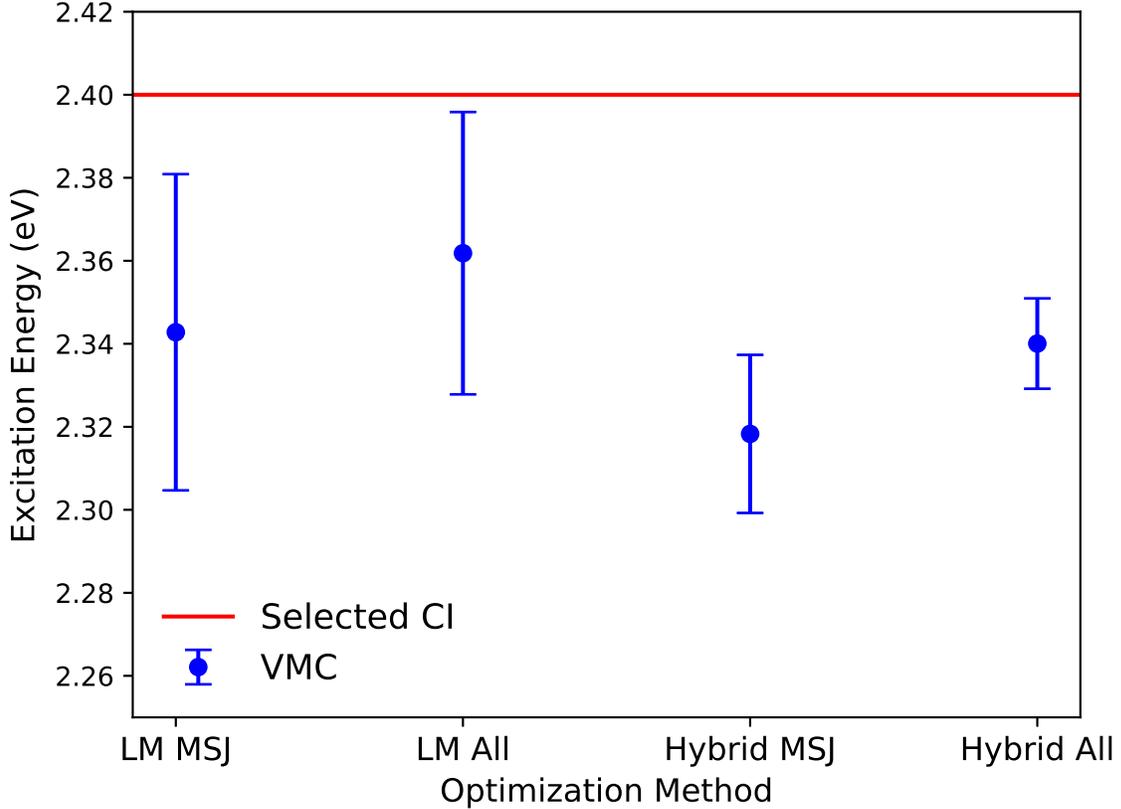}
\caption{Excitation energy of the $2 \; {}^1 \Sigma_g$ state in \ce{C2} for LM and hybrid method on both MSJ and all parameter ansatzes. See also Table \ref{tab:c2_data}. 20 determinants were used for all ground state optimizations. The excited state optimizations used 23 determinants for both the LM and Hybrid MSJ cases, 71 for the LM All case, and 50 for Hybrid All case. The benchmark value is taken from a selected CI calculation using the CIPSI algorithm.\cite{Loos2019}}
\label{fig:c2_results}

\end{figure}

\begin{table}[H]
\caption{Excitation energies and uncertainties for the $ 2 \; {}^1 \Sigma_g$ state in \ce{C2}.
Literature values for selected CI, CCSDT, CC3, and CASPT2 using an aug-cc-pVQZ basis set are included for comparison.\cite{Loos2019}}

\begin{tabular}{ccccc}
Method & \multicolumn{1}{l}{Excitation Energy (eV)}        & \multicolumn{1}{l}{Uncertainty (eV)}  &
\multicolumn{1}{l}{Total Samples}\\ \hline

LM MSJ VMC    & 2.34       & 0.04  & 100,000,000 \\

LM All VMC    & 2.36       & 0.03  & 100,000,000 \\

Hybrid MSJ VMC    & 2.32        & 0.02  &  138,000,000 \\

Hybrid All VMC    & 2.34       & 0.01   &  138,000,000 \\

Selected CI & 2.40 &         &      \\
CCSDT  & 2.87 &         &      \\
CC3  & 3.24 &         &      \\
CASPT2  & 2.50 &         &      \\
\end{tabular}

\label{tab:c2_data}
\end{table}

Figure \ref{fig:c2_results} shows the predicted excitation energy achieved by the LM and 
the hybrid method on both the simpler MSJ ansatzes and on all parameter wave functions that
include the NCJF and orbital rotations.
It is reassuring to find that the hybrid method's results agree with the LM's to within 
statistical uncertainty.
In terms of accuracy, the VMC results are within about 0.05 eV of the selected CI value with 
the all parameter optimizations offering some improvement over the MSJ results.
This has been achieved with very modest CI expansions, using less than 100 determinants 
in all cases, compared to the 5 million used to produce the benchmark energy.\cite{Loos2019}

\begin{figure}[H]
\includegraphics[width=12cm]{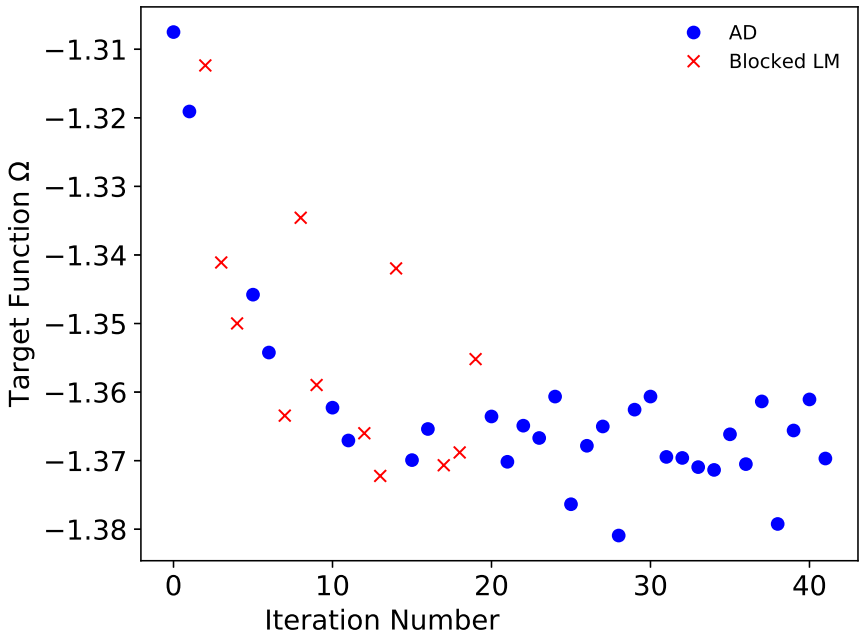}
\caption{Example optimization of target function $\Omega$ for all parameters using the hybrid method. Points for AD correspond to an average over 50 iterations while those for the Blocked LM are individual iterations.}
\label{fig:c2_opt}

\end{figure}

We also explore the potential benefits of the hybrid approach over a pure LM optimization.
Figure \ref{fig:c2_opt} shows an example optimization by the hybrid method of all parameters 
including NCJF and orbitals.
With these more challenging parameters, the Blocked LM is more prone to step uncertainty and
upward fluctuations of the target function, which in this case we see the AD sections correct.
The hybrid method also exhibits a statistical advantage over the LM based on the uncertainty 
and sampling data in Table \ref{tab:c2_data}.
For slightly greater sampling effort, the hybrid method provides about a factor of 3 
improvement in the uncertainty on the excitation energy, which would require a factor of 9 
more samples to obtain solely with the LM.
This advantage in computational efficiency persists in our results for more difficult systems.

\subsection{Nitroxyl, Glyoxal, and Acrolein}

One nuance to the study of doubly excited states is distinguishing between different types of
such states.
For instance, the state we have considered in the carbon dimer is multi-determinantal while 
some other molecules' states can be viewed as single reference, enabling higher order coupled 
cluster\cite{Loos2019} or orbital optimized DFT\cite{Hait2020} to obtain accurate excitation 
energies.
The work of Loos and coworkers has categorized a set of doubly excited states using the 
percentage of singles amplitudes in CC3\cite{Loos2019} and we briefly consider several 
of the same systems to compare our methodology against their benchmark results.
Specifically, we consider the $2 \; {}^1 \text{A}^\prime$ state in nitroxyl, the 
$2 \; {}^1 \text{A}_g$ state in glyoxal, and the $3 \; {}^1 \text{A}^\prime$ state of acrolein.
These systems exhibit some of the diversity of doubly excited states, with 
acrolein's state containing a high percentage of single excitations, while nitroxyl and glyoxal
have almost none.
For all systems, we use cc-pVTZ basis sets with pseudopotentials\cite{Bennett2017} and 
generate determinants for MSJ ansatzes from CASSCF calculations in Molpro.\cite{MOLPRO_brief}
Our active spaces are (12e,9o) for nitroxyl, (8e,6o) for glyoxal, and (10e,10o) for acrolein.
Each CASSCF calculation was state-averaged over four singlet states.

\begin{figure}[H]
\color{red}
\includegraphics{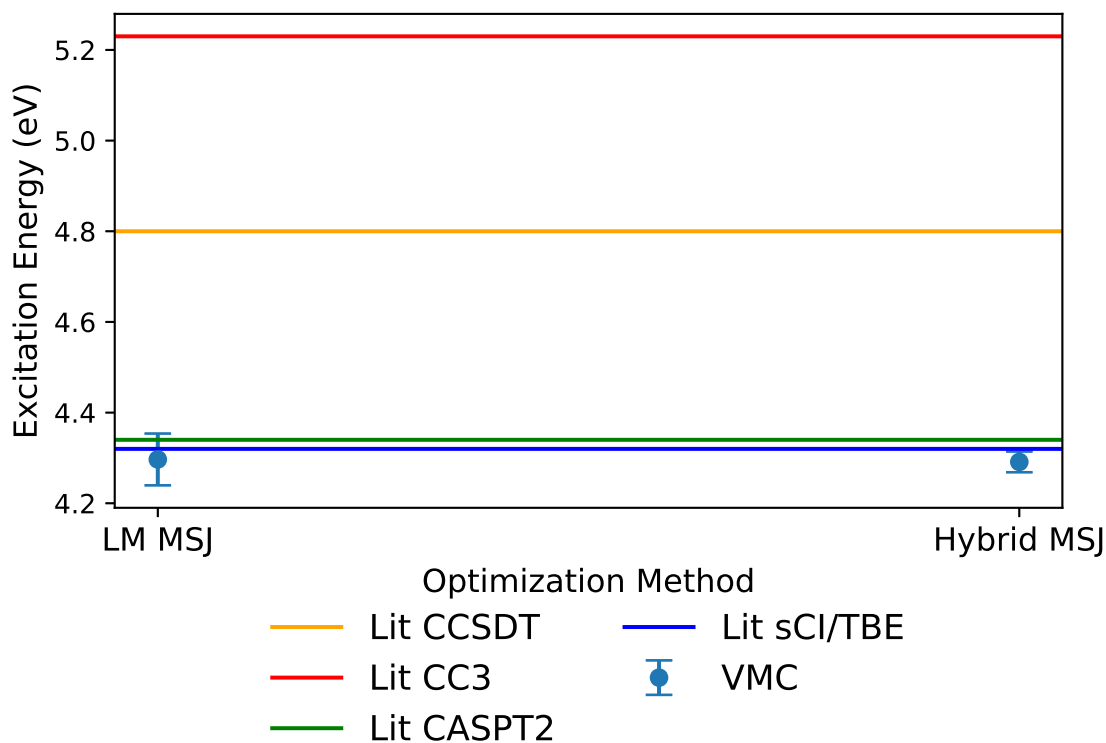}
\caption{Excitation energy of the $2 \; {}^1 \text{A}^\prime$ state in nitroxyl for LM and hybrid method on MSJ ansatzes. See also Table \ref{tab:nitroxyl_data}. Both methods used 20 determinants the ground state and 40 for the excited state. Reference values for coupled cluster, CASPT2, and selected CI are taken from the work of Loos and coworkers.\cite{Loos2019} They take the selected CI value as the theoretical best estimate.}
\label{fig:nitroxyl_results}

\end{figure}

\begin{figure}[H]

\includegraphics{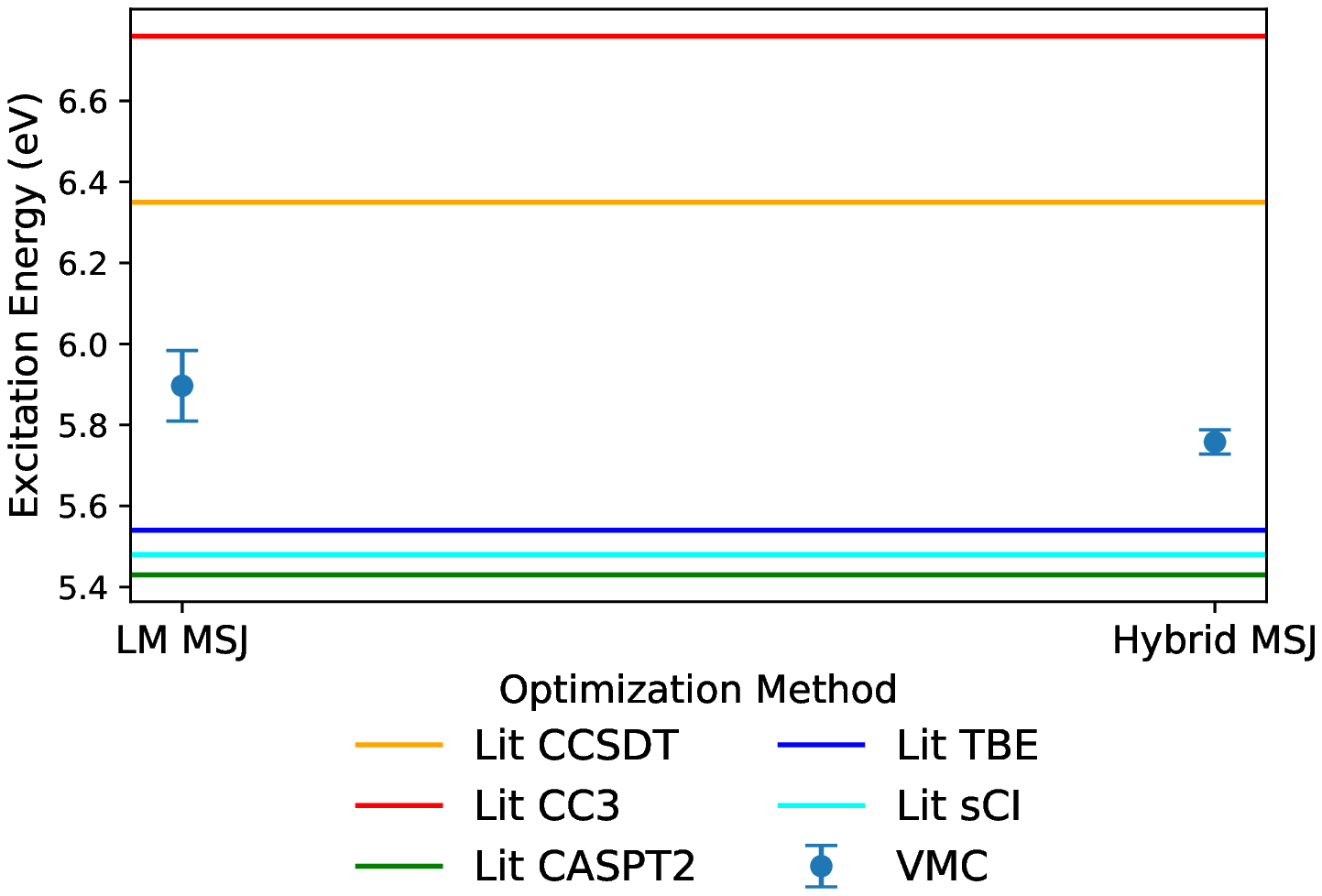}
\caption{Excitation energy of the $2 \; {}^1 \text{A}_g$ state in glyoxal for LM and hybrid method on MSJ ansatzes. See also Table \ref{tab:glyoxal_data}. Both methods used 50 determinants the ground state and 20 for the excited state. Reference values for coupled cluster, CASPT2, and selected CI are taken from the work of Loos and coworkers.\cite{Loos2019} Their theoretical best estimate is obtained by adding a 0.06 eV basis set correction to the selected CI result.}
\label{fig:glyoxal_results}

\end{figure}

\begin{figure}[H]

\includegraphics{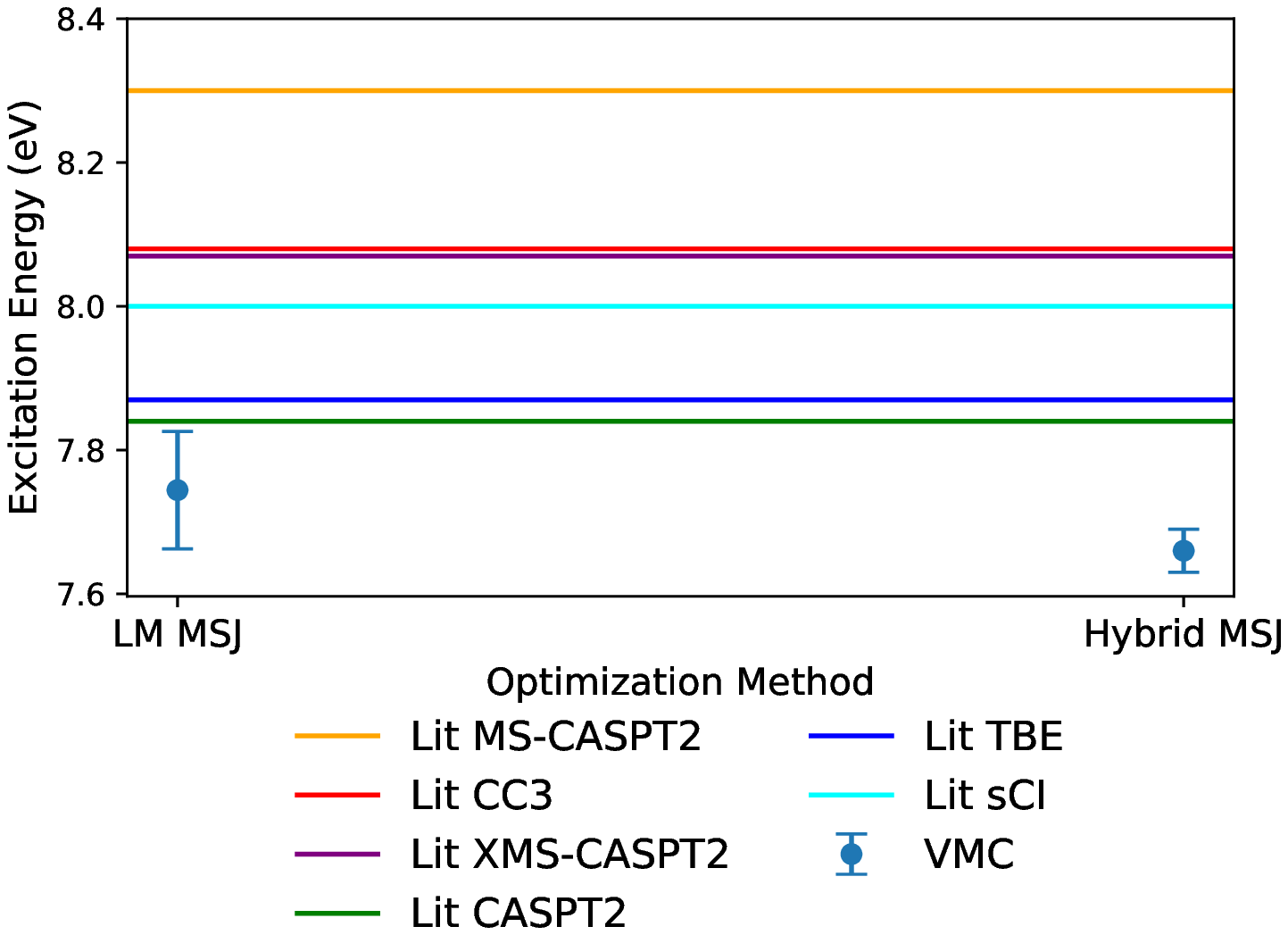}
\caption{Excitation energy of the $3 \; {}^1 \text{A}^\prime$ state in acrolein for LM and hybrid method on MSJ ansatzes. See also Table \ref{tab:acrolein_data}. Both methods used 20 determinants the ground state and 70 for the excited state. Reference values for coupled cluster, CASPT2, and selected CI are taken from the work of Loos and coworkers.\cite{Loos2019} Their theoretical best estimate is obtained by adding a -.13 eV basis set correction to the selected CI result.}
\label{fig:acrolein_results}

\end{figure}

The excitation energy predictions from the LM and the hybrid method are shown in Figures 
\ref{fig:nitroxyl_results} through \ref{fig:acrolein_results} with precise values given in 
Tables \ref{tab:nitroxyl_data} through \ref{tab:acrolein_data}.
We compare our results to the theoretical best estimates (TBE) from Loos and coworkers as well
as their coupled cluster and CASPT2 values.\cite{Loos2019}
For this trio of systems, we find that we can obtain good accuracy with very modest wave 
functions and consistent results between the LM and hybrid method.
For nitroxyl, we come within 0.03 eV of the TBE, while our excitation energies in glyoxal and 
acrolein are within about 0.2 eV.
This level of accuracy outperforms the coupled cluster approaches as well as some versions of 
CASPT2 in the case of acrolein.
As in the carbon dimer, these calculations use less than 100 determinants in all cases
compared to the millions used in the benchmark selected CI calculations, which are 
restricted to smaller basis sets in glyoxal and acrolein.
We note that our methodology allows for further systematic improvement through more 
sophisticated ansatzes and in order to consider its performance on larger systems beyond the 
reach of selected CI, we now turn to some molecules outside the benchmark set of Loos and 
coworkers.

\begin{table}[H]

\caption{Excitation energies and uncertainties for the $2 \; {}^1 \text{A}^\prime$ state in nitroxyl. The included literature values for CCSDT, CC3, CASPT2, and selected CI all use an aug-cc-pVQZ basis set.\cite{Loos2019}}
\begin{tabular}{ccccc}
Method & \multicolumn{1}{l}{Excitation Energy (eV)}  & \multicolumn{1}{l}{Uncertainty (eV)}  &
\multicolumn{1}{l}{Total Samples}\\ \hline

LM MSJ VMC    & 4.30       & 0.06  & 100,000,000 \\
Hybrid MSJ VMC    & 4.29        & 0.01  &  138,000,000 \\
TBE/Selected CI  & 4.32 &         &      \\
CCSDT  &4.8 &         &      \\
CC3  & 5.23 &         &      \\
CASPT2  &4.34 &         &      \\
\end{tabular}

\label{tab:nitroxyl_data}
\end{table}

\begin{table}[H]

\caption{Excitation energies and uncertainties for the $2 \; {}^1 \text{A}_g$ state in glyoxal. The included literature values use an aug-cc-pVDZ basis set for selected CI  and aug-cc-pVQZ for CCSDT, CC3, and CASPT2.\cite{Loos2019}}
\begin{tabular}{ccccc}
Method & \multicolumn{1}{l}{Excitation Energy (eV)}  & \multicolumn{1}{l}{Uncertainty (eV)}  &
\multicolumn{1}{l}{Total Samples}\\ \hline

LM MSJ VMC    & 5.90      & 0.09  & 140,000,000 \\

Hybrid MSJ VMC    & 5.75        & 0.03  &  138,000,000 \\

TBE & 5.54 &         &      \\
Selected CI  & 5.48 &         &      \\
CCSDT  & 6.35 &         &      \\
CC3  & 6.76 &         &      \\
CASPT2  & 5.43 &         &      \\
\end{tabular}

\label{tab:glyoxal_data}
\end{table}

\begin{table}[H]

\caption{Excitation energies and uncertainties for the $3 \; {}^1 \text{A}^\prime$ state in acrolein. The included literature values use a 6-31+G(d) basis set for selected CI, an aug-cc-pVTZ basis CC3, and an aug-cc-pVQZ basis for the three versions of CASPT2.\cite{Loos2019}}
\begin{tabular}{ccccc}
Method & \multicolumn{1}{l}{Excitation Energy (eV)}  & \multicolumn{1}{l}{Uncertainty (eV)}  &
\multicolumn{1}{l}{Total Samples}\\ \hline

LM MSJ VMC    & 7.74      & 0.08  & 140,000,000 \\

Hybrid MSJ VMC    &   7.66      &  0.03 &  138,000,000 \\

TBE & 7.87 &         &      \\
Selected CI  & 8.00 &         &      \\
CC3  & 8.08 &         &      \\
CASPT2  & 7.84 &         &      \\
MS-CASPT2  & 8.3 &         &      \\
XMS-CASPT2  & 7.84 &         &      \\
\end{tabular}

\label{tab:acrolein_data}
\end{table}

\subsection{Cyclopentadiene}
For a more challenging test of our methodology, we consider the doubly excited $3 \; {}^1 
A_1$ state of cyclopentadiene (CPD).
This state has been repeatedly studied in theoretical benchmark 
investigations\cite{Watts1996,Schreiber2008,Silva-Junior2008,Shen2009,Silva-Junior2010,Piotr2015} and in some experimental investigations\cite{Frueholz1979,McDiarmid1985}.
As before, we construct multi-Slater wave functions and add traditional 1 and 2-body 
Jastrow factors.
For this larger molecule, we use the heatbath selected CI (HCI) 
method\cite{Holmes2016,Sharma2017} in the Dice code to produce our CI expansions by correlating
26 electrons in the lowest 46 orbitals from a (6e,5o) CASSCF in Molpro with 
pseudopotentials and cc-pVTZ basis sets.\cite{Bennett2017}

\begin{figure}[H]
\includegraphics{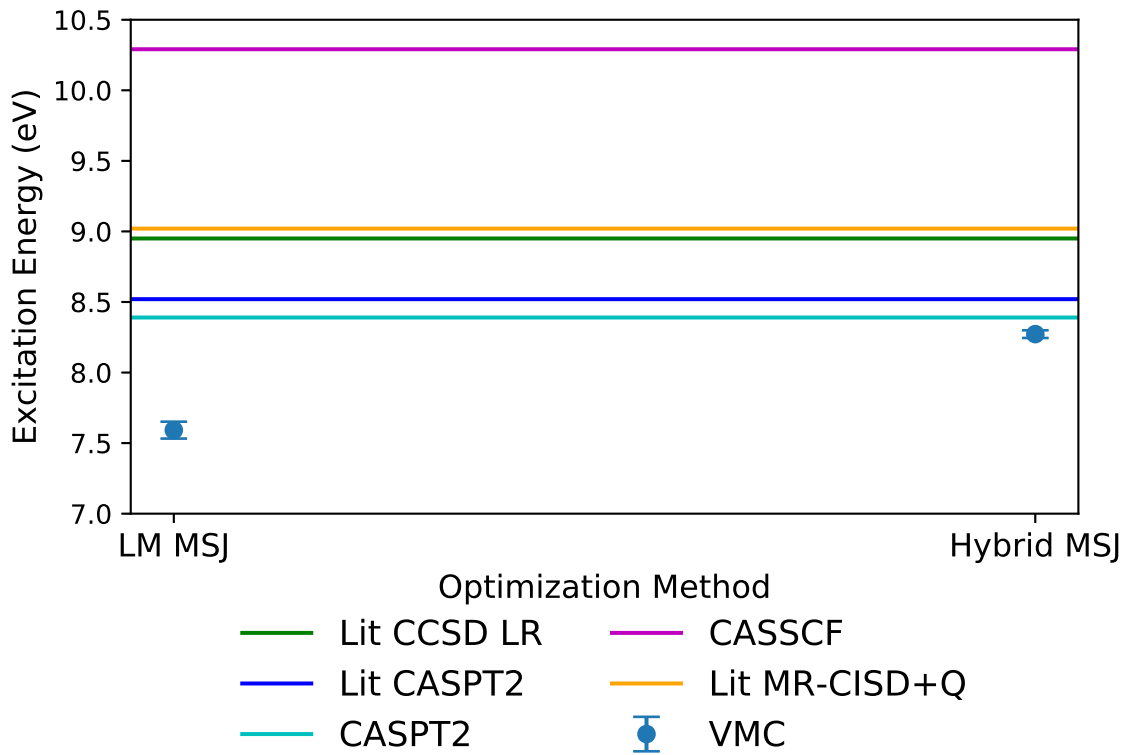}
\caption{Excitation energy of $3 \; {}^1 A_1$ state in CPD for LM and hybrid method. See also 
Table \ref{tab:cpd_data}. Both the LM and hybrid optimizations used 20 and 500
determinants for the ground and excited states, respectively.
Literature values are taken from Schreiber 
et al.\cite{Schreiber2008} for CASPT2 and CCSD LR and from Shen and Li\cite{Shen2009} for MR-CISD+Q.}
\label{fig:cpd_results}

\end{figure}

In this system, we find that the LM fails to optimize the $\Omega$ functional as well as the
hybrid method and leads to an inferior energy prediction.
However, for relatively simple variance-matched multi-Slater Jastrow ansatzes 
of 20 and 500 determinants for the ground and excited states respectively, the hybrid method is
able to achieve an excitation energy within about 0.1 to 0.2 eV of CASPT2 as seen in 
Figure \ref{fig:cpd_results} and Table \ref{tab:cpd_data}.

\begin{table}[H]
\caption{Excitation Energies and uncertainties for $3 \; {}^1 
A_1$ state in CPD.}

\begin{tabular}{ccccc}
Method & \multicolumn{1}{l}{Excitation Energy (eV)}        & \multicolumn{1}{l}{Uncertainty (eV)}  &  
\multicolumn{1}{l}{Total Samples}\\ \hline

LM MSJ VMC    & 7.59       & 0.06   & 140,000,000\\ 

Hybrid MSJ VMC    & 8.27        & 0.03    &  210,000,000\\
CASSCF          & 10.29        &       \\
CASPT2          & 8.39 &      &  \\
Lit. CASPT2\cite{Schreiber2008}          & 8.52 &      &  \\
Lit. CCSD LR\cite{Schreiber2008}          & 8.95 &      &  \\
Lit. MR-CISD+Q\cite{Shen2009}          & 9.02 &      &  \\

\end{tabular}

\label{tab:cpd_data}
\end{table}

To elaborate on the failure of the LM in this case, we present data in Table 
\ref{tab:cpd_compare_data} on a head to head comparison of the LM and the hybrid method on 
the same excited state wave function with a determinant expansion length of 50 and using the
same fixed value of $\omega$ in the target function.
We find that the hybrid method achieves a lower value of the target function, 
which corresponds to a significantly higher energy for the excited state.
This results in its substantially better prediction for the excitation energy as both 
optimization methods give comparable results for the ground state calculation.
We note that extensive experimentation with the technical details of the LM, 
including choices for the shifts, sampling effort, and guiding function failed to bring it into
agreement with the hybrid method.
When we started the LM with the hybrid method's \emph{optimized} wave function, it remained 
at roughly those parameter values with the same target function value and energy as found 
by the hybrid method.
This test indicates that the LM agrees that the hybrid method has found an optimal location in 
parameter space if it starts close enough, but is apparently unable to find it itself when 
starting from the unoptimized wave function.

\begin{table}[H]
\caption{Head to head comparison of LM and hybrid for CPD on the same excited state 
wave function at fixed $\omega$.}

\footnotesize
\begin{tabular}{cccccc}
Method & \multicolumn{1}{|p{2.5cm}|}{Energy (a.u.)}    & \multicolumn{1}{|p{3cm}|}{Energy Uncertainty (a.u.)} &
\multicolumn{1}{|p{2.5cm}|}{Target Function $\Omega(\Psi)$ (a.u.)} & \multicolumn{1}{|p{3cm}|}{Target Uncertainty (a.u.)} \\
\hline

LM MSJ VMC    & -31.541       & 0.0013    & -0.747 & 0.0023 \\ 
Hybrid MSJ VMC & -31.517 &      0.0007  & -0.752 & 0.0004

\end{tabular}

\label{tab:cpd_compare_data}
\end{table}

\subsection{4-Aminobenzonitrile}
Our final system, 4-aminobenzonitrile (ABN), has been heavily studied as an example of 
intramolecular charge transfer(ICT) with many attempts to determine the geometry of the ICT
state\cite{Gomez2005,Segado2016a,Segado2016b,Tran2020}.
Here we test our ability to treat a doubly excited state at the ICT geometry.
We selected this system for an initial exploration of possible benefits of using 
state-specific orbitals within VMC while forgoing orbital optimization.
These orbitals were obtained from a recent state-specific CASSCF 
approach\cite{Tran2019,Tran2020} that employs a root tracker based on combination of an excited
state variational principle and density matrices.
For a twisted geometry of ABN\cite{Tran2020} (coordinates in supplementary material), we 
construct Multi-Slater wave functions from both 
state-averaged (over four states) and state-specific CASSCF calculations that use a (12e,11o) active space along
with pseudpotentials and cc-pVDZ basis sets.\cite{Bennett2017}
Both types of CASSCF calculations were performed in a development version of 
PySCF.\cite{Sun2018}
The excited state with double excitation character that we consider appears as the fourth 
CASSCF state in energy, directly above the ICT state.

\begin{figure}[H]
\includegraphics{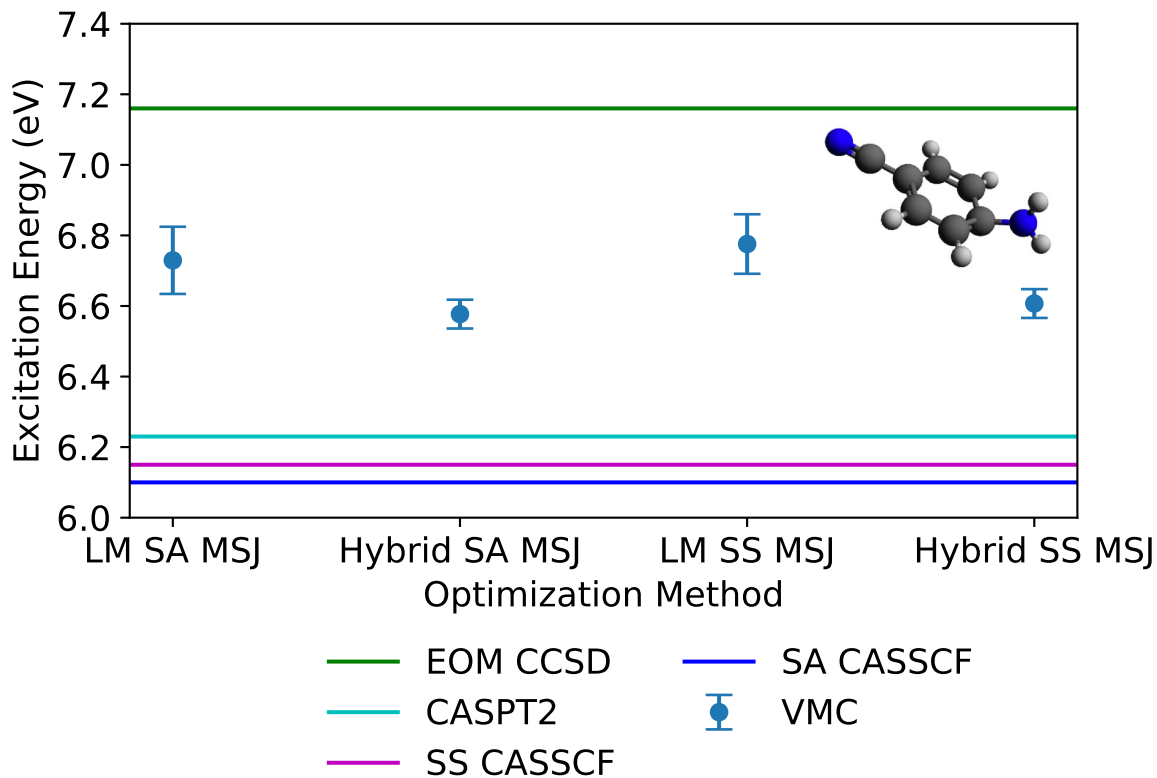}
\caption{Excitation energy of the doubly excited state in ABN for LM and hybrid method. See also Table \ref{tab:abn_data}. Excitation energies for all other methods were obtained in Molpro. 20 determinants were used in all ground state optimizations. The excited state optimizations used 70, 100, 500, and 300 determinants for the LM SA, Hybrid SA, LM SS, and Hybrid SS cases respectively. These numbers of determinants were chosen to achieve an explicit variance match between the ground and excited states for each method.}
\label{fig:abn_results}

\end{figure}

Figure \ref{fig:abn_results} shows the excitation energies obtained by the LM and hybrid method
for the cases where we use state-averaged and state-specific CASSCF orbitals.
In this instance, we find that the optimization methods agree with each other and that there is
no clear difference between using state-averaged and state-specific orbitals within our VMC
ansatzes.
There is about a 0.4 eV difference between our VMC results and CASPT2, but in the absence of an
experimental result or higher level benchmark, it is not obvious which is more accurate.
The agreement between the state-averaged and state-specific VMC excitation energies may not be 
too surprising given that we also find little difference at the CASSCF level.
While this is a null result for the usefulness of state-specific orbitals for this state in 
ABN, other cases may perform differently, including the ICT state.
In terms of optimization, the across the board agreement in ABN offers further evidence that 
the hybrid method is at least as accurate as the LM, while continuing to provide better 
statistical efficiency.

\begin{table}[H]
\caption{Excitation Energies and uncertainties for ABN.}

\begin{tabular}{ccccc}
Method & \multicolumn{1}{l}{Excitation Energy (eV)}        & \multicolumn{1}{l}{Uncertainty (eV)}   
& \multicolumn{1}{l}{Total Samples}\\ \hline

LM SA MSJ VMC    & 6.73      & 0.10 & 140,000,000 \\

Hybrid SA MSJ VMC    & 6.58         & 0.04   & 174,000,000 \\
LM SS MSJ VMC    & 6.78      & 0.08  &  140,000,000\\

Hybrid SS MSJ VMC    & 6.61        & 0.04  &   192,000,000 \\
SA CASSCF          & 6.10        &     &  \\
SS CASSCF          & 6.15 &    &   \\
SA CASPT2           & 6.23 &     &    \\
EOM-CCSD            & 7.16 &    &

\end{tabular}

\label{tab:abn_data}
\end{table}

\section{Conclusion}

We have presented an extension of a hybrid LM/AD optimization
approach to the case of excited-state-specific variance
optimization and tested its efficacy for doubly excited
states.
As in energy minimization, we find the hybrid method to be
more statistically efficient than the linear method,
and, in one case, we were surprised to find it to be
more effective at finding the variance minimum.
Thanks to VMC's ability to combine a linear combination
of determinants (for capturing strong correlation)
with sophisticated correlation factors (for weak correlation)
and its ability to explicitly balance wave function quality
between different states, we find it to be highly
accurate compared to theoretical benchmarks
in our tests on doubly excited states.
As it relies on far more modest determinant expansions than
sCI methods, it can also be used to treat both strong and
weak correlation in system sizes where capturing both
through sCI is not currently possible.
We have also performed 
some simple tests
on the stability of state-specific variance minimization
and found that using symmetry, increasing sample sizes,
and improving the quality of the wave function approximation all
play important roles in preventing collapse to other states.

%In this work, we have first provided some degree of comfort on the stability of 
%variance-based state-specific VMC optimization.
%While Filippi and coworkers have raised important concerns and we can certainly encounter 
%pitfalls in practice, practitioners of excited state VMC can still obtain robust results with 
%sufficient experience and care.
%The hybrid method has been shown to be a useful tool in this endeavor, obtaining excitation 
%energies of comparable or even significantly superior accuracy to the LM and with greater 
%statistical efficiency.
%More generally comparing to other electronic structure methods, we have also found that our VMC
%methodology is capable of achieving highly accurate results for doubly excited states using 
%relatively modest MSJ ansatzes.
%The opportunity for systematic improvement through more flexible ansatzes and further
%optimization effort, which the development of the hybrid method supports, allows VMC to play an
%important benchmarking role for these states that challenge many other quantum chemistry 
%methods.

Looking forward, there are multiple avenues for further improvement.
Our optimization stability testing was limited to a relatively simple
example, and while it does suggest steps that can be taken
to alleviate stability concerns, it remains to be seen
how effective these steps will be in general.
The demonstration by Filippi and coworkers \cite{Cuzzocrea2020}
that optimization instabilities can persist even when
highly sophisticated CIPSI-based VMC trial functions are
employed makes clear that there is more work to be done
to resolve this issue.
As variance-based approaches are particularly ill-suited to degenerate
or near-degenerate states, it would be quite interesting to explore
whether generalized variational principles that incorporate properties
beyond the energy can be usefully adapted for VMC optimization.
Another priority is improving user-accessibility, as the recent
improvements in VMC optimization methodology have in many cases
brought with them a significant increase in the methodological
complexity.
Finding ways to robustly automate choices for stability shifts,
what balance to strike between linear method and accelerated
descent steps, what hyperparameters to choose for the descent,
and how to arrange variables into blocks within the blocked
linear method would significantly simplify the practical
application of these tools.
The systematic study of effective importance sampling functions
is another supporting step for excited state optimizations, where their use is 
more crucial than in the ground state.
A third improvement would be to automatically stage the optimization of 
different parameters according to the statistical significance of their gradients, which would
allow the noisiest and most difficult parameters to be handled at the end of the optimization
without any direct human decision-making. 
Finally, although VMC optimization is becoming increasingly
capable, it will likely be profitable to map out areas
where difficult parameters like orbital shapes can be safely
kept at their quantum chemistry values, whether from
state-averaged or state-specific CASSCF.
Given the high accuracy that VMC can offer for very challenging
excited states, providing easy-to-use incarnations of the
best available VMC optimization methods is a high priority.

\section*{Supplementary Material}

%\begin{suppinfo}

See the supplementary material for molecular geometries, absolute energies, and optimization details for doubly excited states.

%\end{suppinfo}

\section*{Data Availability}

The data that support the findings of this study are available from the corresponding author upon reasonable request.
Input and output files for QMCPACK are available via the Materials Data Facility with the DOI: https://doi.org/10.18126/kduf-y6cp 
\begin{acknowledgement}

This work was supported by the U.S. Department of Energy, Office of Science, Basic Energy Sciences, Materials Sciences and Engineering Division, as part of the Computational Materials Sciences Program and Center for Predictive Simulation of Functional Materials.
Computational work was shared evenly between the Berkeley Research Computing
Savio cluster and the LBNL Lawrencium cluster.

\end{acknowledgement}


\bibliography{references}

\end{document}

\section{ Molecular Geometries }
  
  \begin{table}[H]
%\small
%\footnotesize
     \caption{Structure of CN5. Coordinates in \r{A}.}
     \centering
     \begin{tabular}{lSSS}
       \multicolumn{1}{l}{} &  &  & \\ \hline
 C     &     0.000000   &  0.000000   &  0.340620 \\
C      &    0.000000    & 1.188120   &  -0.363814 \\
C      &    0.000000 &   -1.188120     &  -0.363814 \\
H       &   0.000000  &  0.000000    & 1.424530 \\
H        &  0.000000   &  1.160200     &  -1.449150 \\
H        &  0.000000  &  -1.160200    &  -1.449150 \\
N         & 0.000000   &  2.389120    &  0.164413 \\
N   &      0.000000   & -2.389120    &  0.164413 \\
H    &      0.000000   & 2.531430    &  1.161940 \\
H     &     0.000000   &  -2.531430    &  1.161940 \\
H      &    0.000000   & 3.211380    & -0.414923 \\
H       &   0.000000   &  -3.211380    &  -0.414923 \\
       \hline
     \end{tabular}
     \label{tab:cn5Geo}
   \end{table}
  
      \begin{table}[H]
      %\color{red}

     \caption{Structure of nitroxyl. Coordinates in \r{A}.}
     \centering
     \begin{tabular}{lSSS}
       \multicolumn{1}{l}{} &  &  & \\ \hline
O     &    0.111655    &  0.000000   &  1.140178 \\
N      &    -0.236949    & 0.000000   &  -0.018994 \\
H      &    0.625294 &   0.000000     &  -0.621185\\

       \hline
     \end{tabular}
     \label{tab:cpdGeo}
   \end{table}
   
       \begin{table}[H]
       %\color{red}

     \caption{Structure of glyoxal. Coordinates in \r{A}.}
     \centering
     \begin{tabular}{lSSS}
       \multicolumn{1}{l}{} &  &  & \\ \hline
 C     &     0.642211    &  0.401329   &  0.000000 \\
C      &    -0.642211    & -0.401329   &  0.000000 \\
O      &    1.722903 &   -0.139984    &  0.000000\\
O       &   -1.722903  &  0.139984    & 0.000000 \\
H        &  0.508726   &  1.491662     & 0.000000 \\
H        &  -0.508726  &  -1.491662    &  0.000000 \\
       \hline
     \end{tabular}
     \label{tab:cpdGeo}
   \end{table}

       \begin{table}[H]
       %\color{red}

     \caption{Structure of acrolein. Coordinates in \r{A}.}
     \centering
     \begin{tabular}{lSSS}
       \multicolumn{1}{l}{} &  &  & \\ \hline
C     &     -0.590800    &  -0.361686   & 0.000000 \\
C      &    0.638441    & 0.442999   &  0.000000 \\
C      &   1.835351 &   -0.152787     &  0.000000\\
O       &   -1.712769  &  0.101534    & 0.000000 \\
H        &  -0.426590   &  -1.453900     & 0.000000 \\
H        &  0.522297  &  1.516693    &  0.000000 \\
H         & 2.756647   &  0.409814    &  0.000000 \\
H   &      1.910074   & -1.232987    &  0.000000 \\

       \hline
     \end{tabular}
     \label{tab:cpdGeo}
   \end{table}

    \begin{table}[H]

     \caption{Structure of cyclopentadiene. Coordinates in \r{A}.}
     \centering
     \begin{tabular}{lSSS}
       \multicolumn{1}{l}{} &  &  & \\ \hline
 H     &     -0.879859    &  0.000000   &  1.874608 \\
H      &    0.879859    & 0.000000   &  1.874608 \\
H      &    0.000000 &   2.211693     &  0.612518\\
H       &   0.000000  &  -2.211693    & 0.612518 \\
H        &  0.000000   &  1.349811     & -1.886050 \\
H        &  0.000000  &  -1.349811    &  -1.886050 \\
C         & 0.000000   &  0.000000    &  1.215652 \\
C   &      0.000000   & -1.177731    &  0.285415 \\
C    &      0.000000   & 1.177731    &  0.285415 \\
C     &     0.000000   &  -0.732372    &  -0.993420 \\
C      &    0.000000   & 0.732372    & -0.993420 \\
       \hline
     \end{tabular}
     \label{tab:cpdGeo}
   \end{table}
   
       \begin{table}[H]

     \caption{Structure of 4-aminobenzonitrile. Coordinates in \r{A}.}
     \centering
     \begin{tabular}{lSSS}
       \multicolumn{1}{l}{} &  &  & \\ \hline
   C  & 0.031570    & -0.001678     &  1.801388 \\
   C  & 0.046873    & -1.225280     & 1.069477 \\
   C  & 0.035483    & -1.235146     & -0.304583 \\
   C  &-0.115039    & 0.000908      & -1.046464 \\
   C  & 0.031512    & 1.235978      & -0.302121 \\
   C   &0.042951    &  1.223413     &   1.071925 \\
   C  & 0.047593    & -0.003141     & 3.231727 \\
   H  & 0.052819    & -2.162843     & 1.609066 \\
   H  & 0.044455    & -2.181757     & -0.831777 \\
   H  & 0.037413    &  2.183658     & -0.827425 \\
   H  & 0.045866    &  2.159919     & 1.613382 \\
   N  & 0.059850    & -0.006281     & 4.396999 \\
   N  & 0.660760    &  0.003409     & -2.284953 \\
   H  & 0.225584    &  0.003579     & -3.196767 \\
   H  & 1.677891    &  0.005163     & -2.289493 \\
       \hline
     \end{tabular}
     \label{tab:abnGeo}
   \end{table}
   
 \section{Absolute Energies and Optimization Details for Doubly Excited States}
 
 %\textcolor{red}{
 Our reported excitation energies in the main text are the energy differences 
 between ground and excited state optimizations of approximately matching variances, and the 
 total numbers of samples we report are the sum of the amounts used for the two optimizations.
 We note that obtaining variance matched values in practice requires optimizations at 
 multiple determinant expansion lengths beyond pairs of ground and excited state results 
 presented here.
 For our double excitations results, all LM optimizations used our adaptive three shift scheme 
 described in Section 2.2 of the main text with 1,000,000 samples per iteration.
 Hybrid optimization consisted of multiple macro-iterations each composed of 100 AD iterations 
 and 3 Blocked LM iterations before a descent finalizing section of 1100 iterations.
 In all cases over all systems, each Blocked LM iteration used 1,000,000 samples, which were 
 doubled in our calculation of the total sampling effort as the Blocked LM runs over its sample 
 twice and each AD iteration used 30,000 samples, both within the hybrid method and the pure
 descent finalizing section.
 Energies and uncertainties for each state were computed from an average over the last 10 steps 
 for LM optimizations and from the last 500 descent finalizing steps for hybrid
 optimizations.
 %}
 
 We took a staged approach to our excited state optimizations by first optimizing only a 
 short CI expansion alongside the one- and two-body Jastrow factors, and then adding 
 additional determinants until the variance matched that of the ground state.
 In the cases where they were used, NCJF and orbital parameters were only optimized starting 
 from a pre-optimized MSJ wave function.
 In the hybrid method, AD sections were used to identify 5 vectors in parameter space that 
 were provided to the Blocked LM steps.
 %\textcolor{red}{
 The Blocked LM sections divided parameters into 5 blocks in all hybrid 
 optimizations. 
 As discussed in section 2.2 of the main text, the Blocked LM first performs a LM-style matrix 
 diagonalization for each block and retains some of the eigenvectors.
 For the carbon dimer MSJ case, nitroxyl, glyoxal, and acrolein, 10 eigenvectors 
 were retained from each block and in all other cases, 30 eigenvectors 
 were retained.
 In each hybrid optimization, these retained eigenvectors were combined with the 5 vectors from 
 the preceding AD section for another diagonalization to obtain the Blocked LM's parameter 
 update.
 %}
 % Note I'm commenting out the red color for the submitted version
 % for the supp material.  For the main text I will submit both versions.
 On AD sections, we lowered initial step sizes for the later optimization stages to avoid  
 kicking the optimization out of the target function minimum.
 Table \ref{tab:initialStepSizes} shows the initial step sizes used for different parameter
 types when they were unoptimized.
 For later stages, the step sizes for the Jastrow factors and CI coefficients were reduced 
 by factors of 10 and 2 respectively.
 The NCJF and orbital parameters required relatively small changes in their values compared 
 to the other parameter types and we found that significantly smaller step sizes for them were 
 effective in stably improving the target function.
 %\textcolor{red}{
 Based on our experience from this work, these choices for step sizes and 
 Blocked LM parameters within our staged optimizations form a reasonable default option for
 applying our methodology in similar studies.
 However, adjusting these settings may be necessary in challenging cases to either achieve a 
 better optimization or at least verify that one is not easily obtainable.
 %}
   \begin{table}[H]
%\footnotesize
   \caption{Initial step sizes for completely unoptimized parameters of different types.}
    \centering
    \begin{tabular}{lcc}
         Parameter Type & Step Size
      \\ \hline
         One-Body Jastrow & 0.1  \\
         Two-Body Jastrow & 0.1  \\
         CI Coefficient   & 0.01  \\
         NCJF F Matrix       & 0.0005 \\
         Orbital Parameter & 0.0005 \\
      
    \end{tabular}
    \label{tab:initialStepSizes}
\end{table}
   
%\textcolor{red}{
In tables \ref{tab:c2_energies} through \ref{tab:abn_energies}, we list the 
total number of optimizable parameters and the absolute energies for each of the 
systems we consider in our doubly excited state calculations.
The parameter numbers are for results with CI expansion lengths chosen to obtain an explicit 
variance match between the ground and excited state.
In some cases, the LM and hybrid method needed different expansion lengths to variance match 
their respective ground state results leading to different total numbers of excited state 
parameters.
Our statistical uncertainties on the last digit of the absolute energies are shown in parentheses.
In table \ref{tab:all_iterations}, we list the number of iterations in each optimization for the 
different cases in all our molecules.
%}

  \begin{table}[H]
  %\color{red}
%\footnotesize
   \caption{Absolute energies and parameter numbers for the carbon dimer.}
    \centering
    \begin{tabular}{lcccc}
    & GS Parameters & ES Parameters & GS Energy (a.u.) & ES Energy (a.u.) 
      \\ \hline
          LM MSJ & 50 &  53 & -11.0232(7) & -10.9372(7) \\
      Hybrid MSJ & 50 &  53 & -11.0230(3) & -10.9377(4)  \\
      LM All & 1080 &  1251 & -11.0338(6) & -10.9469(6) \\
      Hybrid All & 1080 &  1170 & -11.0343(2) & -10.9483(1)  \\
      
    \end{tabular}
    \label{tab:c2_energies}
\end{table}

\begin{table}[H]
  %\color{red}
%\footnotesize
   \caption{Absolute energies and parameter numbers for nitroxyl.}
    \centering
    \begin{tabular}{lcccc}
    & GS Parameters & ES Parameters & GS Energy (a.u.) & ES Energy (a.u.) 
      \\ \hline
          LM MSJ & 70 &  90 & -26.387(1) & -26.229(1) \\
      Hybrid MSJ & 70 &  90 & -26.3858(4) & -26.3851(4)  \\

    \end{tabular}
    \label{tab:nitroxyl_energies}
\end{table}

\begin{table}[H]
  %\color{red}
%\footnotesize
   \caption{Absolute energies and parameter numbers for glyoxal.}
    \centering
    \begin{tabular}{lcccc}
    & GS Parameters & ES Parameters & GS Energy (a.u.) & ES Energy (a.u.) 
      \\ \hline
          LM MSJ & 70 &  90 & -44.445(2) & -44.228(2) \\
      Hybrid MSJ & 70 &  90 & -44.4440(6) &-44.2324(5)  \\

    \end{tabular}
    \label{tab:nitroxyl_energies}
\end{table}

\begin{table}[H]
  %\color{red}
%\footnotesize
   \caption{Absolute energies and parameter numbers for acrolein.}
    \centering
    \begin{tabular}{lcccc}
    & GS Parameters & ES Parameters & GS Energy (a.u.) & ES Energy (a.u.) 
      \\ \hline
          LM MSJ & 70 &  120 & -35.319(2) & -35.319(1) \\
      Hybrid MSJ & 70 &  120 & -35.3169(6) &-35.3169(5)  \\

    \end{tabular}
    \label{tab:nitroxyl_energies}
\end{table}

  \begin{table}[H]
  %\color{red}
%\footnotesize
   \caption{Absolute energies and parameter numbers for cyclopentadiene.}
    \centering
    \begin{tabular}{lcccc}
          & GS Parameters & ES Parameters & GS Energy (a.u.) & ES Energy (a.u.) 
      \\ \hline
          LM MSJ & 60 &  540 & -31.830(1) & -31.551(1)  \\
      Hybrid MSJ & 60 &  540 & -31.8287(4) & -31.5253(6)  \\

    \end{tabular}
    \label{tab:cpd_energies}
\end{table}

  \begin{table}[H]
  %\color{red}
%\footnotesize
   \caption{Absolute energies and parameter numbers for 4-aminobenzonitrile.}
    \centering
    \begin{tabular}{lcccc}
          & GS Parameters & ES Parameters & GS Energy (a.u.) & ES Energy (a.u.) 
      \\ \hline
          LM SA MSJ & 70 &  120 & -62.855(2) & -62.608(2)  \\
           Hybrid SA MSJ & 70 &  150 & -62.8453(8) & -62.6037(6)  \\
      LM SS MSJ & 70 &  550 & -62.855(2) & -62.606(2)  \\
      Hybrid SS MSJ & 70 &  350 & -62.8463(9) & -62.6035(6)  \\
      
    \end{tabular}
    \label{tab:abn_energies}
\end{table}

  \begin{table}[H]
  %\color{red}
%\footnotesize
   \caption{Iteration numbers for all systems. The hybrid method values are the number of macro-iterations before the descent finalizing section.}
    \centering
    \begin{tabular}{lcc}
    & GS Iterations & ES Iterations
      \\ \hline
        LM MSJ C2 & 50 & 50\\
      Hybrid MSJ C2 &  4 & 4 \\
      LM All C2    &  50 & 50\\
      Hybrid All C2 & 4 & 4 \\
      LM MSJ nitroxyl & 50 & 50 \\
      Hybrid MSJ nitroxyl &4  & 4 \\
      LM MSJ glyoxal &70  &  70\\
      Hybrid MSJ glyoxal &4 & 4\\
      LM MSJ acrolein & 70 & 70 \\
      Hybrid MSJ acrolein & 4 & 4\\
      LM MSJ CPD & 50 & 90 \\
      Hybrid MSJ CPD & 8 & 8 \\
      LM SA MSJ ABN & 70 & 70 \\
       Hybrid SA MSJ ABN & 7 & 5 \\
    LM SS MSJ ABN & 70 & 70 \\
    Hybrid SS MSJ ABN & 7 & 7 \\

    \end{tabular}
    \label{tab:all_iterations}
\end{table}